# Crude oil market and geopolitical events: an analysis based on information-theory-based quantifiers

**Aurelio F. Bariviera**
Department of Business, Universitat Rovira I Virgili. Av. Universitat 1, 43204 Reus, Spain.
aurelio.fernandez@urv.cat

**Luciano Zunino**
Centro de Investigaciones Ópticas (CONICET La Plata-CIC), C.C. 3, 1897 Gonnet, Argentina
Departamento de Ciencias Básicas, Facultad de Ingeniería, Universidad Nacional de La Plata (UNLP), 1900 La Plata, Argentina
lucianoz@ciop.unlp.edu.ar

**Osvaldo A. Rosso**
Instituto de Física, Universidade Federal de Alagoas, BR 104 Norte km 97, 57072-970 Maceió, Alagoas, Brazil
Instituto Tecnológico de Buenos Aires (ITBA), Av. Eduardo Madero 399, C1106ACD Ciudad Autónoma de Buenos Aires, Argentina
Complex Systems Group, Facultad de Ingeniería y Ciencias Aplicadas, Universidad de los Andes, Av. Mons. Álvaro del Portillo 12.455, Las Condes, Santiago, Chile
oarosso@gmail.com

**Abstract**

This paper analyzes the informational efficiency of oil market during the last three decades, and examines changes in informational efficiency with major geopolitical events, such as terrorist attacks, financial crisis and other important events. The series under study is the daily prices of West Texas Intermediate (WTI) in USD/BBL, commonly used as a benchmark in oil pricing. The analysis is performed using information-theory-derived quantifiers, namely permutation entropy and permutation statistical complexity. These metrics allow capturing the hidden structure in the market dynamics, and allow discriminating different degrees of informational efficiency. We find that some geopolitical events impact on the underlying dynamical structure of the market.

## 1. Introduction

It is needless to say that crude oil represents a strategic asset and a very important commodity for the world economy. Since the oil embargo in the 1970s, the market experienced several important disruptions, due to global economic or political events. There are many studies in the economic literature which address very different research questions. Brahmasrene et al. (2014) studies the causality between crude oil prices and foreign exchange rates. An et al. (2014) examines the relationship between spot and future prices of crude oil. Sun and Shi (2015) look for structural breaks in prices time series, scrutinizing the presence of unit roots, as a fingerprint for a nonstationary process. Wang et al. (2015) focus their interest in price forecasting.

It has been reported that some economic variables, such as liquidity (Bariviera, 2011) or an extended financial crisis (Bariviera et al., 2012), affect the levels of informational efficiency in stock and bond markets. Our interest in this paper is to study the informational efficiency of the crude oil market, and we aim to discriminate different dynamical regimes that arise contemporary of several important geopolitical events.

The concept of informational efficiency is the kingpin in financial economics. Fama (1970) defines it as a situation in the market when prices reflect all available information. Ross (2004) recognizes that it is the consequence of the interaction of a large set of individual decisions made on some underlying information. As a corollary, with the same information set it is not possible to obtain superior returns. It implies, also, that future returns depend to a great extent not only on historic information but also on the new information that arrives at the market.

The detailed analysis of time series could be the starting point for the description and understanding of the system under analysis. One of the usual topics in economics is to uncover the underlying dynamics that governs the evolution of a given time series. Within economics, finance is one of the richest fields to apply advanced time series methods. The reason for this is that financial markets produce a great amount of information, recorded even at tiny intervals. Consequently, large datasets are readily available for analysis.

The aim of this paper is to assess the temporal evolution of the informational efficiency of the crude oil market, by analyzing the time series of one of the main benchmarks: West Texas Intermediate prices. The rest of the paper is organized as follows. Section 2 describes the information-theory-based methodology. Section 3 details the data under analysis and presents the results. Finally, Section 4 draws the main conclusions from our study.

**2. Information theory quantifiers**

When studying real-world phenomena, it is usual to record measurements of these phenomena manifestations. If these measurements are taken at different times, their sequence is commonly known as time series. When the data are abundant, the number of suitable techniques for their treatment increases. In particular, econophysics methods, as the one applied in this article, are innovative and appropriate to shed light on economic problems. In many cases, econophysics overcomes the limitations of traditional econometric techniques. In this line, information-theory-derived quantifiers can help to extract relevant information from financial time series.

During the 1960s, there was a series of papers that used entropy in order to study the predictability of time series. Theil and Leenders (1965) used entropy to predict short-term price fluctuations in the Amsterdam Stock Exchange. This original use was followed by Fama (1965) and Dryden (1968), who replicate the study for the New York Stock Exchange and the London Stock Exchange, respectively. Hart (1971) studies several measures of income concentration, including some entropy-inspired measures. Philippatos and Wilson (1974) propose the average mutual information or shared entropy as a proxy of systematic risk. Although information theory techniques were somewhat forgotten in economic analysis, it was reintroduced by econophysicists in recent years. For example, Risso (2008) uses entropy and symbolic time series analysis in order to relate informational efficiency and the probability of having an economic crash. Later, Risso (2009) implements Shannon entropy to rank the informational efficiency of several stock markets around the world. Rosso et al. (2010) highlight the potentiality of information theory related approaches in for unveiling nonlinear dynamics.

**2.1 Entropy**

One of the measures selected in our paper is the Shannon entropy (Shannon and Weaver, 1949). Given a discrete probability distribution P = { $p_i$ : i = 1 , . . . M} , Shannon entropy is defined as: $S[P] = -\sum_{i=1}^{M} p_i \log p_i$ . This formula measures the information embedded into the physical process described by the probability distribution P. It is a bounded function in the interval [0, log (M)]. If S[P]=0 the process exhibits one state with probability equal to one and the remaining states equal to zero. In other words, zero entropy means full certainty about the system's outcome. On the other extreme, if S[P]= log (M) = $S_{max}$, our knowledge about the system is minimal, meaning that all states are equally probable.

**2.2 Statistical complexity**

Even though entropy can describe globally the level of order/disorder of a process, the analysis of time series using solely Shannon entropy could be incomplete (Feldman and Crutchfield, 1998; Feldman et al., 2008). The reason is that an entropy measure does not quantify the degree of structure or patterns present in a process.

An additional measure that quantifies the hidden structure of the process is needed to fully characterize dynamical systems: a statistical complexity measure. Two extreme situations, perfect order and maximum randomness (e.g., a periodic sequence and a fair coin toss) reveal no structure. A complexity measure should vanish in such cases. Between these two situations, there is a variety of processes with different degrees of physical structure.

A family of statistical complexity measures, based on the functional form developed by López-Ruiz et al. (1995) is defined in Lamberti et al. (2004) and Martín et al. (2006) as: $C_{JS}[P] = Q_J[P, P_e] \cdot H[P]$, where $H[P] = \frac{S[P]}{S_{max}}$ is the normalized Shannon entropy, P is the discrete probability distribution of the time series under analysis, $P_e$ is the uniform distribution and $Q_J[P,P_e]$ is the so-called disequilibrium. This disequilibrium is defined in terms of the Jensen–Shannon divergence, which quantifies the difference between two probability distributions. Martín et al. (2006) demonstrate the existence of upper and lower bounds for generalized statistical complexity measures such as $C_{JS}$. This complexity measure is able to capture essential details of the dynamics and distinguish different degrees of periodicity and chaos. Additionally, as highlighted in Zunino et al. (2010b), the statistical complexity is not a trivial function of the normalized entropy because it is based on two probability distributions.

**2.3 Estimation of the probability density function**

In order to evaluate these quantifiers, we should previously estimate the probability density function (PDF). There are several methods to do it. Beyond the traditional histogram made by frequency counting, PDF could be estimated by Fourier analysis (Powell et al.,1979), binary symbolic dynamics (Mischaikow et al., 1999), wavelets transformations (Rosso et al., 2001), amplitude-based procedures (De Micco et al., 2008), among many other techniques.

Following Rosso et al. (2007), Zunino et al. (2010a,2011) and Bariviera et al. (2015,2016) we use the Bandt and Pompe (2002) permutation method, because it is the simplest symbolization technique that considers time causality. This methodology requires only weak stationarity assumptions. The appropriate symbol sequence arises naturally from the time series. "Partitions" are devised by comparing the order of neighboring relative values rather than by apportioning amplitudes according to different levels. No model assumption is needed because Bandt and Pompe (BP) method makes partitions of the time series and orders values within each partition. Given a time series $S(t) = \{x_t; t=1,..., N\}$, an embedding dimension $D>1$, $D \in \mathbb{N}$, and an embedding delay $\tau$, $\tau \in \mathbb{N}$, the BP-pattern of order D generated by

$$s \rightarrow (x_s, x_{s+\tau}, \ldots, x_{s+(D-2)\tau}, x_{s+(D-1)\tau}) \quad [1]$$

is the one to be considered. To each time s, BP assign a D-dimensional vector that results from the evaluation of the time series at times s, s+τ,…,s+(D −2) τ, s +(D −1) τ. Naturally, the value of D, reflects the information about "the past" incorporated into the vectors. By the ordinal pattern of order D related to the time s, BP mean the permutation $\pi = (r_0, r_1, \cdots, r_{D-1})$ of $(0, 1, \cdots, D-2, D-1)$ defined by $x_{s+r_0\tau} \leq x_{s+r_1\tau} \leq \ldots x_{s+r_{D-2}\tau} \leq x_{s+r_{D-1}\tau}$. In this way the vector defined by Equation [1] is converted into a definite symbol π. So as to get a unique result BP consider that $r_i < r_{i+1}$ if $x_{s+r_i\tau} = x_{s+r_{i+1}\tau}$. This is justified if the values of $x_t$ have a continuous distribution so that equal values are very unusual. For all the D! possible orderings (permutations) $\pi_i$ when embedding dimension is D, their associated relative frequencies can be naturally computed according to the number of times this particular order sequence is found in the time series, divided by the total number of sequences. Thus, an ordinal pattern probability distribution $P=\{p(\pi_i), i=1, \cdots, D!\}$ is obtained from the time series. As we mention previously, the ordinal-pattern's associated PDF is invariant with respect to nonlinear monotonous transformations. Accordingly, nonlinear drifts or scaling artificially introduced by a measurement device will not modify the quantifiers' estimation, a nice property if one deals with experimental data (see i.e. Saco et al., 2010). These advantages make the BP approach more convenient than conventional methods based on range partitioning. Additional advantages of the method reside in (i) its simplicity (we need few parameters: the pattern length/embedding dimension D and the embedding delay τ) and (ii) the extremely fast nature of the pertinent calculation-process. The BP methodology can be applied not only to time series representative of low dimensional dynamical systems but also to any type of time series (regular, chaotic, noisy, or reality based).

**2.4 The complexity-entropy causality plane**

For quantifying the informational efficiency of the oil market we use the complexity-entropy causality plane, i.e. a Cartesian plane with the permutation entropy of the time series in the horizontal axis and a so-defined permutation statistical complexity measure in the vertical one. Rosso et al. (2007) showed that this plane is not only useful to identify different stochastic processes, but also to discriminate chaotic dynamics. Additionally, it was successfully used to discriminate the linear and nonlinear correlations present in stocks (Zunino et al., 2010a), corporate bonds (Zunino et al., 2012), commodities (Zunino et al., 2011) and interest rates (Bariviera et al., 2016). The location in the complexity-entropy causality plane allows quantifying the inefficiency of the system under analysis because the presence of temporal patterns derives in deviations from the ideal position associated to a totally random process.

**3. Data and empirical results**

The data for this study are daily prices of crude oil WTI FOB in USD per barrel, from 10/01/1983 to 11/11/2015, with a total of 8568 data points. All data were retrieved from DataStream.

Using the above data and the methodology described in Section 2, we compute the permutation entropy and permutation statistical complexity for the daily prices of crude oil using sliding windows. The rationale behind the use of a moving sample was that it enables us to study the evolution of these quantifiers during the period under examination. Our sliding windows contains N=300 data points, the frequency is daily (τ=1), and the step between each window is δ=20 data points. In this way, we obtained 413 estimation periods. We believe that with a pattern length D=4 and 300 data points we are able to capture the dynamics of the series under analysis. Each window spans approximately one year and moves one (trading) month ahead. In this way, changes in the underlying dynamics are taken into account. We obtained similar results using D=3. Taking into account that a larger D value includes more information "of the past", in the input ordinal pattern, we show only the results for D=4.

Figure 1 shows the complexity-entropy causality plane. Each point reflects the calculations of permutation entropy and permutation complexity for a particular window. In order to observe how the informational efficiency changes across time, we grouped windows in 20 groups of 20 windows and one group of 13 windows. These groups are depicted with different markers and colors in order to facilitate the visual inspection.

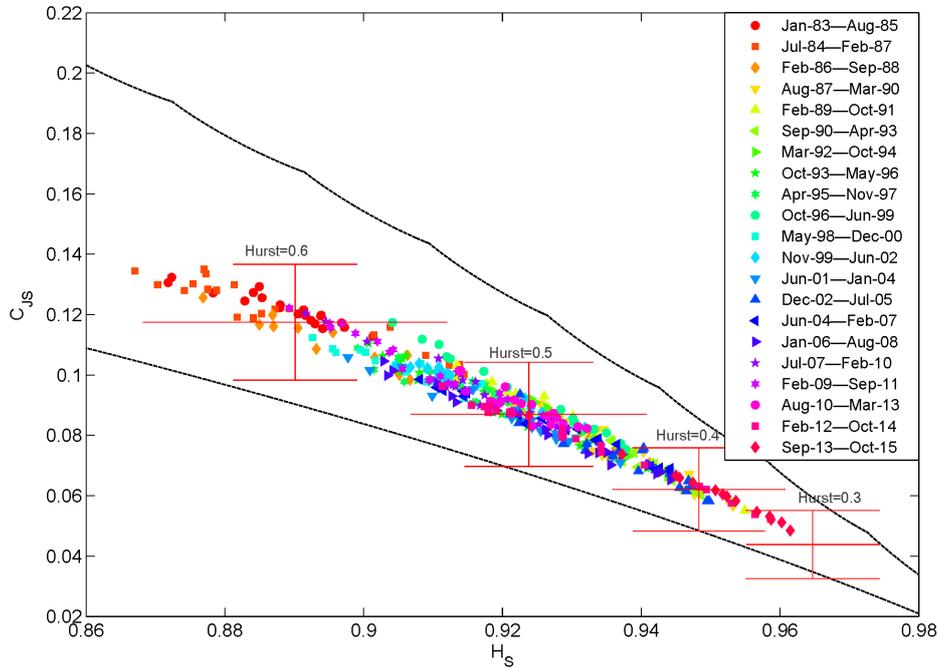

**Figure 1.** Complexity-entropy causality plane for D = 4, $\tau$ = 1, of WTI crude oil daily price. Solid red lines represent the mean and standard deviations values of the quantifiers for fBm simulations for different Hurst exponents. Solid black lines represent the upper and lower bounds of the quantifiers as computed by Martín et al. (2006).

In general, we can say that the time series under study is quite informational efficient (H>0.86 and $C_{JS}$<0.14), since they lie, to a great extent, within the area compatible with a standard Brownian motion (Hurst=0.5). However, different planar locations are associated with different stochastic dynamics. We simulate 100 fractional Brownian motion (fBm) time-series of the same length as the original one and compute the quantifiers using the sliding window procedure and with the same parameters. The simulation was done using the wfbm function from Matlab, for various Hurst exponents (Hurst={0.3,0.4,0.5,0.6}). Then, the average and standard deviation of the simulations were computed for each Hurst value. The average value and its standard deviation are displayed in Figure 1 with solid red lines. As a result of this operation, we observe different underlying dynamics in the long period under examination.

According to the standard financial paradigm, prices in a competitive market should behave randomly. However, this behavior is found only in part of the time series. Specifically, this memoryless behavior is recorded in the points around the zone of Hurst=0.5. Figure 1 shows points reflecting persistent (Hurst>0.5) and antipersistent (Hurst<0.5) behavior. Thus, there are some forces that could affect the memory endowment of time series.

In order to highlight that the planar location of the quantifiers was not obtained by chance, we display in Figure 2 the results of the original series and those of the randomized series. When we shuffle data and thus, destroy all non-trivial correlations, permutation information quantifiers move close to the (1, 0) corner. This picture shows that the proposed quantifiers capture the hidden correlation structures of data.

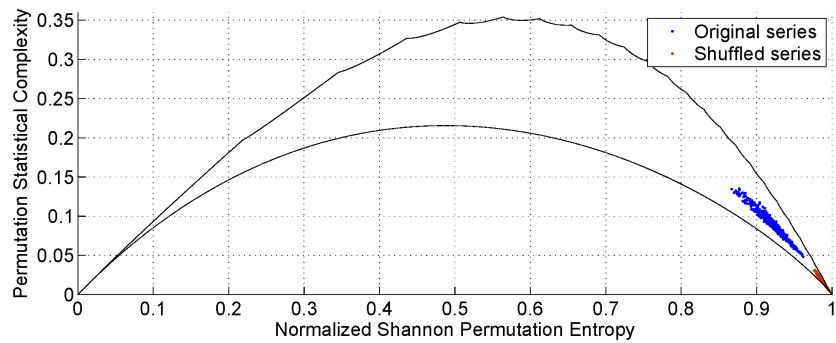

**Figure 2**. Complexity-entropy causality plane, with quantifiers computed for the original and the shuffled time series, for D = 4, τ = 1. The solid black lines represent the upper and lower bounds of the quantifiers as computed by Martín et al. (2006).

It is well known that oil prices are sensitive to geopolitical events. Such events may disrupt current supply or increase uncertainty about future oil supplies. We obtained information from geopolitical events related to the crude oil market from the U.S. Energy Information Administration (http://www.eia.gov/). According to its website, this federal "agency collects, analyzes, and disseminates independent and impartial energy information to promote sound policymaking, efficient markets, and public understanding of energy and its interaction with the economy and the environment."

The major geopolitical and economic events considered in this paper are:

- Saudis abandon swing producer role
- Iraq invades Kuwait
- Asian financial crisis
- OPEC cuts production targets 1.7 mmbpd
- 9-11 attacks
- Low spare capacity
- Global financial collapse
- OPEC cuts production targets 4.2 mmbpd

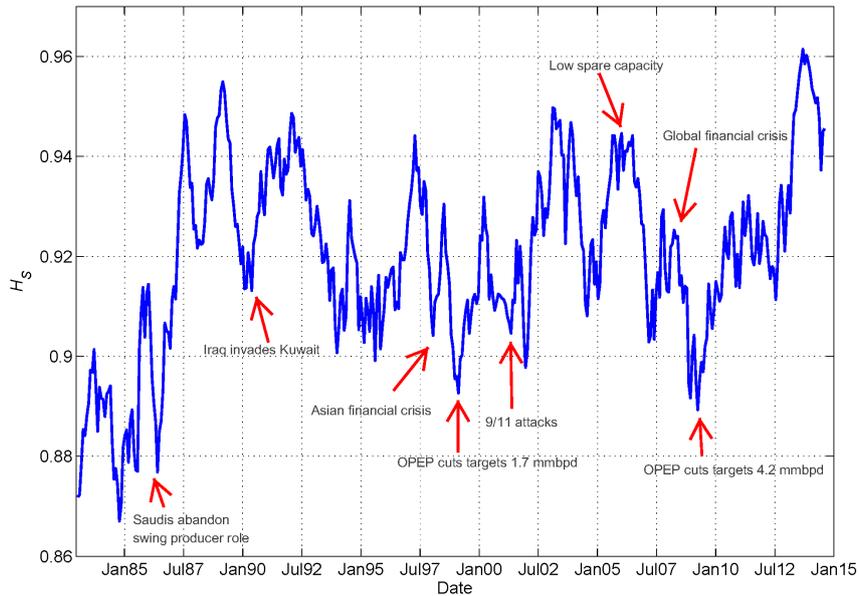

**Figure 3.** Permutation entropy evolution, computed with D=4, $\tau$=1 of the crude oil prices. Geopolitical events were extracted from http://www.eia.gov/finance/markets/crudeoil/spot_prices.php

We can observe in Figure 3, that major geopolitical events are accompanied by great swings in the entropy of the time series. The direction of the movement is not clear. There are some events that trigger up the permutation entropy and others that reduce it. But in both cases, the informational efficiency is not neutral to main political or economic milestones.

**4 Conclusions**

We analyzed the daily values of crude oil WTI FOB in USD per barrel, from 10/01/1983 to 11/11/2015. We assess the informational efficiency of this market, by using information theory derived quantifiers. Even though the crude oil market seems quite efficient, the informational efficiency changed along the 32 years under scrutiny. During this long period, different economic events affected the global economy, and particularly the oil market. We also showed how several events occurred contemporary to changes in the informational efficiency, providing evidence of some influence of main economic and political milestones in the dynamics of the crude oil market.

**References**

An, H., Gao, X., Fang, W., Ding, Y., Zhong, W. (2014). Research on patterns in the fluctuation of the co-movement between crude oil futures and spot prices: A complex network approach. *Applied Energy*, Vol. 136, p. 1067–1075.

Bandt, C.,Pompe, B., 2002. Permutation Entropy: A Natural Complexity Measure for Time Series. *Phys. Rev. Lett.*, Vol. 88, No. 17, p.174102.

Bariviera, A.F. (2011). The influence of liquidity on informational efficiency: The case of the Thai Stock Market. *Physica A: Statistical Mechanics and Its Applications*, Vol. 390, No. 23–24, 4426–4432.

Bariviera, A.F., Guercio, M.B.,Martinez, L. B. (2012). A comparative analysis of the informational efficiency of the fixed income market in seven European countries. *Economics Letters*, Vol. 116, No. 3, p. 426–428.

Bariviera, A.F., Guercio, M.B., Martinez, L. B., Rosso, O.A. (2015). A permutation Information Theory tour through different interest rate maturities: the Libor case. *Philosophical Transactions of the Royal Society of London. Series A*, Vol. 373, No. 2056, 20150119

Bariviera, A.F., Guercio, M.B., Martinez, L.B., Rosso, O.A. (2016). Libor at crossroads: Stochastic switching detection using information theory quantifiers. *Chaos, Solitons & Fractals*, Vol. 88, p. 172-182.

Brahmasrene, T., Huang, J.-C., Sissoko, Y. (2014). Crude oil prices and exchange rates: Causality, variance decomposition and impulse response. *Energy Economics*, Vol. 44, p. 407–412.

De Micco L., Gonzalez C.M., Larrondo H.A., Martín M.T., Plastino A., Rosso O.A. (2008) Randomizing nonlinear maps via symbolic dynamics. Physica A 387, 3373–3383.

Dryden, M.M., (1968). Short-Term Forecasting Of Share Prices: An Information Theory Approach. *Scottish Journal of Political Economy*, Vol. 15, No. 1, p.227–249.

Fama, E.F. 1965. Tomorrow on the New York Stock Exchange. *The Journal of Business*, Vol. 38, No. 3, p.285–299.

Fama, E. F. (1970). Efficient Capital Markets: A Review of Theory and Empirical Work. *The Journal of Finance*, Vol 25, No 2,p. 383–417.

Feldman, D.P., Crutchfield, J.P. (1998). Measures of statistical complexity: Why? *Physics Letters A*, Vol. 238, No. 4–5, 244.

Feldman, D.P., McTague, C.S., Crutchfield, J.P. (2008). The organization of intrinsic computation: Complexity-entropy diagrams and the diversity of natural information processing. *Chaos: An Interdisciplinary Journal of Nonlinear Science*, Vol.18, No. 4, 43106.

Hart, P.E.1971. Entropy and Other Measures of Concentration. *Journal of the Royal Statistical Society. Series A (General)*, Vol. 134, No. 1, p.73–85.

Lamberti, P.W., Martín, M.T., Plastino A., Rosso O.A., Intensive entropic non-triviality measure, 2004 *Physica A: Statistical Mechanics and its Applications* Vol. 334, No. 1-2, 119-131.

López-Ruiz, R, Mancini, H.L., Calbet, X. 1995. A statistical measure ofcomplexity. *Physics Letters A*, Vol. 209, No. 5–6, p. 321–326.

Martín, M.T., Plastino, A., Rosso, O.A. 2006. Generalized statistical complexity measures: Geometrical and analytical properties. *Physica A: Statistical Mechanics and its Applications,* Vol.369, No. 2, p. 439–462.

Mischaikow K, Mrozek M, Reiss J, Szymczak A. 1999 Construction of symbolic dynamics from experimental time series. *Phys. Rev. Lett.*, Vol. 82, p. 1144–1147.

Philippatos, G.C. Wilson, C.J., 1974. Information theory and risk in capital markets. *Omega*, Vol. 2, No. 4, p.523–532.

Powell GE, Percival IC. 1979 A spectral entropy method for distinguishing regular and irregular motion of Hamiltonian systems. *J. Phys. A,* Vol. 12, p. 2053–2071.

Risso, W.A., 2008. The informational efficiency and the financial crashes. *Research in International Business and Finance*, Vol. 22, No. 3, p.396–408.

Risso, W.A., 2009. The informational efficiency: The emerging markets versus the developed markets. *Applied Economics Letters*, Vol. 16, No. 5, p.485–487.

Ross, S.A. 2004. *Neoclassical Finance*, Princeton University Press, Princeton (NJ).

Rosso O.A., Blanco S, Jordanova J, Kolev V, Figliola A, Schürmann M, Basar E. 2001 Wavelet entropy: a new tool for analysis of short duration brain electrical signals. *J. Neurosci. Methods* Vol. 105, p. 65–75.

Rosso, O.A., Larrondo, H.A., Martin, M.T., Plastino, A., Fuentes, M. A. (2007). Distinguishing Noise from Chaos. *Phys. Rev. Lett.*, Vol. 99, No. 15, 154102.

Rosso, O.A., De Micco, L., Plastino, A., Larrondo, H.A. (2010). Info-quantifiers' map-characterization revisited. *Physica A: Statistical Mechanics and Its Applications*, Vol. 389, No. 21, p. 4604–4612.

Saco, P.M., Carpi, L.C., Figliola, A., Serrano, E., Rosso, O.A. (2010). Entropy analysis of the dynamics of El Niño/Southern Oscillation during the Holocene. Physica A: Statistical Mechanics and Its Applications, 389(21), 5022–5027.

Shannon, C.E., Weaver W.(1949)*The Mathematical Theory of Communication* Champaign (IL):University of Illinois Press.

Sun, J., Shi, W. (2015). Breaks, trends, and unit roots in spot prices for crude oil and petroleum products. *Energy Economics*, Vol. 50, p. 169–177.

Theil, H. Leenders, C.T., 1965. Tomorrow on the Amsterdam Stock Exchange. *The Journal of Business*, Vol. 38, No. 3, p.277–284.

US Energy Information Administration. 2016. Crude oil prices and key geopolitical and economic events. Available at: http://www.eia.gov/finance/markets/crudeoil/spot_prices.php [accessed March 5, 2016]

Wang, Y., Liu, L., Diao, X., Wu, C. (2015). Forecasting the real prices of crude oil under economic and statistical constraints. *Energy Economics,* Vol. 51, p. 599–608.

Zunino, L., Zanin, M., Tabak, B. M., Pérez, D.G., Rosso, O.A. (2010a). Complexity-entropy causality plane: A useful approach to quantify the stock market inefficiency. *Physica A: Statistical Mechanics and Its Applications,* Vol.389, No. 9, 1891–1901.

Zunino, L., Soriano, M.C., Fischer, I., Rosso, O.A., Mirasso, C.R. (2010b). Permutation-information-theory approach to unveil delay dynamics from time-series analysis.*Phys.Rev.E*, Vol. 82, No. 4, 46212.

Zunino, L., Tabak, B.M., Serinaldi, F., Zanin, M., Pérez, D.G., Rosso, O.A. (2011). Commodity predictability analysis with a permutation information theory approach. *Physica A: Statistical Mechanics and Its Applications*, Vol. 390, No. 5, 876–890.

Zunino, L., Fernández Bariviera, A., Guercio, M.B., Martinez, L.B., Rosso, O. A. (2012). On the efficiency of sovereign bond markets. Physica A: Statistical Mechanics and Its Applications, Vol. 391, No. 18, 4342–4349.